\documentclass[onecolumn]{article}
\usepackage{amsmath,amssymb}

\usepackage{hyperref}
\begin{document}
\title{Reconciliation of Zero-Point and Dark Energies\\
in a Friedman Dust Universe with Einstein's Lambda}
\author{James G. Gilson\quad  j.g.gilson@qmul.ac.uk\thanks{
 School of Mathematical Sciences,
Queen Mary University of London, Mile End Road, London E1 4NS,
United Kingdom.}}
\date{April 23, 2007}
\maketitle
\begin{abstract}
In this paper, it is shown that the cosmological model that was introduced
in a sequence of three earlier papers under the title  {\it A Dust Universe Solution to the Dark Energy Problem\/} can be used to resolve the problem of the great mismatch of numerical values between dark energy from cosmology and zero point energy from quantum theory. It is shown that, if the zero point energies for the cosmic microwave background and for all the rest of the universe that is not cosmic microwave background are introduced into this model as two entities, their separate values appear within this theory in the form of a numerical difference. It is this difference that gives the numerical value for the zero point value of Einstein's dark energy density. Consequently, although the two zero point energies may be large, their difference can give the known small dark energy value from cosmology for dark energy density. Issues relating to interpretation, calculation and measurement associated with this result and an interpretation of dark energy as a measure of polarisation of the vacuum are discussed. In the first appendix to this paper, problems associated with the standard model of cosmology are solved by redefining temperature in the dust universe model. In the second appendix of this paper, an examination of the dark matter problem in relation to a general relativistic generalisation of Newton's inverse square law is undertaken. In the third appendix to this paper, the formalism is used to derive a formula that gives a possible value for the mass of the universe in terms of Newton's gravitation constant, Einstein's Lambda and the velocity of light. All three appendices have their own detailed abstracts.
\vspace{0.4cm}
\vskip 0.5cm 
\centerline{Keywords: Dust Universe, Dark Energy, Friedman Equations,}
\centerline{Entropy, Population Inversion, Negative Temperature, Zero-Point Energy}
\vskip 0.2cm
\centerline{PACS Nos.: 98.80.-k, 98.80.Es, 98.80.Jk, 98.80.Qc}
\end{abstract}
\section{Introduction}
\setcounter{equation}{0}
\label{sec-intr}
The work to be described in this paper is an application of the cosmological model introduced in the papers {\it A Dust Universe Solution to the Dark Energy Problem\/} \cite{45:gil}, {\it Existence of Negative Gravity Material. Identification of Dark Energy\/} \cite{46:gil} and {\it Thermodynamics of a Dust Universe\/} \cite{56:gil}. All of this work and its applications has its origin in the studies of Einstein's general relativity in the Friedman equations context to be found in references (\cite{03:rind},\cite{43:nar},\cite{42:gil},\cite{41:gil},\cite{40:gil},\cite{39:gil},\cite{04:gil},\cite{45:gil}) and similarly motivated work in references (\cite{10:kil},\cite{09:bas},\cite{08:kil},\cite{07:edd},\cite{05:gil}) and 
(\cite{19:gil},\cite{28:dir},\cite{32:gil},\cite{33:mcp},\cite{07:edd},\cite{47:lem},\cite{44:berr}). The applications can be found in 
(\cite{45:gil},\cite{46:gil},\cite{56:gil},\cite{60:gil},\cite{58:gil}\cite{64:gil}). Other useful sources of information are (\cite{3:mis},\cite{44:berr},\cite{53:pap},\cite{49:man},\cite{52:ham},\cite{54:riz}) with the measurement essentials coming from references (\cite{01:kmo},\cite{02:rie},\cite{18:moh},\cite{61:free}).  Further references will be mentioned as necessary.

The application of the cosmological model introduced in the papers $A$ \cite{45:gil}, $B$,\cite{46:gil} and $C$ \cite{56:gil} is to the extensively discussed {\it quantum cosmology mismatch\/} between the numerical values for dark energy and zero point-energy. This issue has invoked very great interest and many papers have been written about it in recent years. Briefly, the problem is that quantum theory systems are usually found to have a non-zero lowest state of energy which is called the zero-point energy of the system. In classically described systems, on the other hand, usually the energy of a physical system can assume the value zero. In much physical system analysis, only changes of energy are important for the description of what is happening and so the zero point energy is no problem. However, when the zero point-energy has to be taken into account the difficulty arises that only in very simple situations can it be calculated with any ease. As the system becomes more complicated the calculation of the zero point energy becomes greatly more difficult or even impossible. Thus the most popular system to use for the analysis of zero point energy is the quantum simple harmonic oscillator of definite frequency $\nu$, say. In this case, the oscillator has a zero point energy given by $h\nu/2$ which is easily obtained and clearly has a finite value. The big issue concerns the suggestion that the Einstein concept of {\it dark energy\/} that has recently been discovered to exist and be measured and thought to be the cause of an acceleration of the expansion of universe is in fact quantum {\it zero point\/} energy. The difficulty in making this connection is that if the measured density of Einsten's dark energy density has a value given by $\rho_\Lambda$ where $\Lambda$ is his cosmological constant and the theoretical quantum calculated value for the density of zero point energy is given by $\rho_{Z}$
then approximately the numerical ratio of these two quantities is given by the outrageous and incomprehensible numerical result, 
\begin{eqnarray}
\rho_{Z}/\rho_\Lambda \approx 10^{120}.\label{q0}
\end{eqnarray}
This is the sort of value obtained by Beck and Mackey by putting the cut-off integration frequency equal to the Planck frequency and also obtained by many researchers using other routs to calculate zero point values.
This seems to put right out of question the possibility the zero point and dark energies could {\it somehow\/} be very closely related or indeed equal.

One serious impediment to the discussion of zero point energy in relation to the cosmos is its extremely complex forms, configurations and uncertain theoretical conditions. The cosmos is very far from simple and it is not known how its zero point energy can be calculated. However, calculating zero point energy in quantum theory although also usually extremely difficult is fairly simple for a few cases. In particular there are some definite ideas about how it should be calculated in the case of the electromagnetic field and this is all that will be needed to carry through the project of this paper. A detailed analysis of the general question of the relation between zero point energy and dark energy can be found in the recent paper by Beck and Mackey (\cite{57:bec}). That paper gives a comprehensive account of the dark energy zero point energy problem with many references to origins and useful related work. In particular, that paper devotes much discussion as to how dark energy might be measured in the laboratory. From that paper,
 I first copy one of their basic equations to use as a starting point for the work in this paper. From quantum statistical mechanics and quantum field theory one can write down for the electromagnetic field case, here denoted by $\Gamma$, a number per unit volume of oscillators at temperature $T$ and frequency $\nu$ as follows:

\begin{eqnarray}
\rho_\Gamma(\nu,T)= \left\{\frac{4\pi\nu^3}{c^3}\right\} \left\{1 + \frac{2}{  \exp(\frac{h\nu}{kT}) -1 }\right\}.\label{q1}
\end{eqnarray}

Apart from the quantum simple harmonic oscillator itself, this is the simplest 
example of a formula giving a specific form for the quantum zero point energy, the energy at $T=0$, of a physical system that can be {\it used\/} to seriously explore the problems of zero point energy. However, even this formula presents us with deep difficulties. The two terms are represented separately at (\ref{q2}) and (\ref{q3}) and their sum is given at (\ref{q4}). The $z$ in the subscript meaning {\it zero point part\/} and the $b$ meaning {\it non zero point part\/}.
\begin{eqnarray}
\rho _{\Gamma z}(\nu)&=& \frac{4\pi\nu^3}{c^3}  \label{q2}\\
\rho _{\Gamma b} (\nu,T)&=& \left\{\frac{4\pi\nu^3}{c^3}\right\} \frac{2}{\exp(\frac{h\nu}{kT}) -1}\label{q3}\\
\rho_\Gamma (\nu,T)&=& \rho _{\Gamma z}(\nu) + \rho _{\Gamma b}(\nu,T).\label{q4}
\end{eqnarray}
The first of these difficulties arises if we wish to use these equations to explore a system that involves an enclosure containing all possible frequencies within the electro magnetic field. In that case, the integration of $\rho _{\Gamma z}(\nu)$ from zero to infinity with respect to $\nu$ gives from the first term, the zero point energy in (\ref{q5}), an infinite amount of mass equivalent of the energy per unit volume,
\begin{eqnarray}
\lim _{\nu_c\rightarrow\infty} \int_0^{\nu _c} \rho _{\Gamma z}(\nu)h d\nu &=& \infty .\label{q5}
\end{eqnarray}
where $\nu _c$ is the Beck-Mackey (\cite{57:bec}) cut-off frequency that, if kept finite, avoids the infinity. If the zero point energy of any system we chose to study does actually turn out to be infinite then we are in great trouble with interpretation or being able to use the theory at all. This, of course, is why Beck and Mackey (\cite{57:bec}) introduce the cut off and indeed suggest the theory may well require some future modification so that this difficulty can be bypassed. However, if you attempt to follow through the various and numerous ways, through statistics, quantum, classical, or thermodynamics that the formula (\ref{q4}) has been obtained with many dubious filling in steps, it does become clear that  the formula can only be said to be a {\it very good tentative\/} approximation to the physical truth. Thus if we attempt to apply this formula as it stands to the cosmological context its deficient and approximate nature will likely confuse the issue of whether zero point energy and dark energy are coincident concepts. Thus I take the optimistic view that the correct or at least a better version of the formula will one day be found and anticipate its properties as being those aspects that are possessed by equation (\ref{q4}) less the adverse characteristics. Of course, I cannot give a detailed formula to replace equation (\ref{q4}) but fortunately that is not necessary. It should be remarked that the greatest weight of these difficulties is centred on the zero point energy component $ \rho _{\Gamma z}(\nu)$ of (\ref{q4}). The $\rho _{\Gamma b}(\nu,T) $  component is much safer and that can be integrated from $0$ to $\infty$ with respect to $\nu$ to give the usual fourth power of $T$ formula for energy density of black body radiation,
\begin{eqnarray}
\rho_{\Gamma B}= \int_0^{\infty} \rho _{\Gamma b} (\nu,T)h d\nu  =  aT^4.\label{q6}
\end{eqnarray}
Thus to apply this set of ideas to the cosmological context I define the following density and integral formulas and their properties. I shall use the same symbols and notation but the {\it explicit\/} forms (\ref{q2}, \ref{q3}, \ref{q4}) will not from now on be consider to apply. The integral cut off frequency will also be retained but not assumed to be equal to the Planck frequency. From now on it will be assumed that the
$\lim \nu _c \rightarrow \infty$ applied to the upper limit of an integral gives a {\it finite\/} but possibly large answer. I have allowed for a possible time dependence in the dark energy density components but it is possible that this is not necessary and all such contributions will usually be {\it not\/} variable with respect to time. It is possible that exceptions could occur to this constancy with respect to time near the singularity at $t=0$.  
\begin{eqnarray}
\rho_{\Gamma Z,\nu _c}(t) &=& \int_0^{ \nu _c} \rho _{\Gamma z} (\nu,t) h d\nu \label{q7}\\
\rho_{\Gamma B,\nu _c}(T) &=& \int_0^{\nu _c} \rho _{\Gamma b }(\nu,T) h d\nu\label{q8}\\
\rho _{\Gamma, \nu _c}(t) &=& \rho_{\Gamma Z,\nu _c}(t) + \rho_{\Gamma B,\nu _c}(T).\label{q9}
\end{eqnarray} with the further conditions
\begin{eqnarray}
\lim_{\nu _c \rightarrow \infty }\rho_{\Gamma Z,\nu _c}(t)&=&\ \rho_{\Gamma Z}(t)\  = finite\  function\ of\ time\label{q10}\\
\lim_{\nu _c \rightarrow \infty }\rho_{\Gamma B,\nu _c}(T) &=& aT^4\ =\ 
black\ body\ form\label{q11}
\end{eqnarray}
The time dependent version of $\rho_{\Gamma B,\nu _c}(T)$ to be used in the rest of this paper is defined through the assumed time dependence of the temperature $T(t)$ as
\begin{eqnarray}
\rho_{\Gamma B,\nu _c}(t) = \rho_{\Gamma B,\nu _c}(T(t)) \approx aT^4(t).\label{q11.1}
\end{eqnarray}
The last seven equations involve the symbol $\rho$ as representing energy density as a result of the integration of $\rho$ as a number density with respect to $\nu_c$ and the multiplier h.
The application of this set of quantities to the cosmological dark energy problem is now straight forward and essentially amounts to replacing quantities in the original theory with a form that includes their zero point energy terms. The problem is: can this be done consistently with the relation between dark energy and zero point energy becoming understandable and not involving the massive numerical mismatch that has seemed to be involved?  
\section{Dark-Energy and Zero Point Energy}
\setcounter{equation}{0}
\label{sec-dezp}
The cosmological model introduced in references A, B, C like most cosmological models is a highly idealised version of what physical reality actually involves. This idealisation process is immediately noticeable in the character of the basic mass density $\rho (t)$ for what I call the total conserved mass, $M_U$, of the universe. This is conserved in the sense that $M_U$ is an absolute constant unchanging over the whole life history of the model from $t=-\infty$ through a singularity at $t=0$ and on to $t=+\infty$. While the density is idealised in the sense that $\rho (t)$ only depends on the epoch time, t, and does not vary with position within the universe, $r_i < r(t)$, $r(t)$ being the spherical {\it radius\/} of the universe at epoch, t.
From now on the integrated densities still denoted by the $\rho $ symbol  will represent mass density. In fact they are the energy densities of the last section divided by $c^2$. This is necessary to conform to the cosmological analysis of the previously published work in $A$, $B$  and $C$.  
In $C$, the cosmic microwave background, $CMB$, was introduced into the model by dividing the density $\rho (t)$ into two parts, the contribution from the $CMB$, $\rho_\Gamma$ and the rest as the density $ \rho_\Delta$. 
\begin{eqnarray}
M_U &=& \rho (t)V_U(t) =\ an\ absolute\ constant\label{q13}\\
\rho (t) &=& \rho_\Delta (t) + \rho_\Gamma (t)\label{q14}\\
 M_U &=&(\rho_\Delta (t) + \rho_\Gamma (t))V_U(t)= M_\Delta + M_\Gamma  .\label{q15}
\end{eqnarray}
I also make the {\it strong\/} assumption that the total masses of the two parts $M_\Delta$ and $M_\Gamma$ are both absolute constants. Thus over the history of the system these two parts do not exchange energy or mass. Thus a first move in introducing zero point energies into the model is to replace $M_\Delta$ and $ M_\Gamma $ with their zero point energy versions $M_{\Delta ,\nu_c}$ and $M_{\Gamma ,\nu _c}$, say, making use of the $\nu_c$ parameter to distinguish between the original and changed quantities. Clearly the total universe conserved mass $M_U$ will theoretically change under this operation so that we should now write equation (\ref{q16}) in place of equation (\ref{q15}). However, $M_U$ is an observational input parameter so that it will remain unchanged numerically.
\begin{eqnarray}
 M_{U,\nu_c} = M_{\Delta ,\nu_c} + M_{\Gamma ,\nu_c} .\label{q16}
\end{eqnarray}
I have introduced the Zero point energy case, $ M_{\Delta ,\nu_c}$, for the component of universe mass that is not $CMB$ without any introductory discussion about how it might be calculated from quantum theory. It is obvious that we can have practically no idea how any realistic calculation can be carried through for such a complicated system as what is in fact the greater part of the observable universe  with all its many diverse fields and interactions. However, as remarked earlier, we deal with idealisations which in fact iron out the complications of reality. Thus it is appropriate to introduce the properties for the zero point energy of the $\Delta$ mass field as in (\ref{q7}) to (\ref{q11}) for the $CMB$ field apart from the conversion from energy to mass factor $c^{-2}$,
\begin{eqnarray}
\rho_{\Delta Z,\nu _c}(t) &=& \int_0^{ \nu _c} \rho _{\Delta z} (\nu,t) (h/c^2) d\nu \label{q17}\\
\rho_{\Delta B,\nu _c}(t) &=& \int_0^{\nu _c} \rho _{\Delta b }(\nu,t) (h/c^2) d\nu\label{q18}\\
\rho _{\Delta, \nu _c}(t) &=& \rho_{\Delta Z,\nu _c}(t) + \rho_{\Delta B,\nu _c}(t).\label{q19}
\end{eqnarray} with the further conditions
\begin{eqnarray}
\lim_{\nu _c \rightarrow \infty }\rho_{\Delta Z,\nu _c}(t)&=&\ \rho_{\Delta Z}(t) = finite\  function\ of\ time\label{q20}\\
\lim_{\nu _c \rightarrow \infty }\rho_{\Delta B,\nu} (t) &=&\ non\ zero\ point\ part.\label{q21}
\end{eqnarray}
The main difference between the $\Delta$ functions and the $\Gamma$ functions is that temperature, $T$, is replaced by epoch time, $t$. The capital B or Z subscripts indicate the frequency integrated version of the lower cases b or z density functions. Now that we have some notation, the substitution process can proceed. Firstly, let us consider the mechanical equilibrium equation for pressures obtained in $C$, 
\begin{eqnarray} 
P(t) &=& P_G + P_\Lambda \equiv 0 \ \forall\  t\label{q22}\\
P_G &=& P_\Delta (t)  - P_\Gamma (t) \label{q23}\\
P_G &=& - P_\Lambda  = P_\Delta (t)  - P_\Gamma (t) \label{q24}\\
P_\Lambda  &=& P_\Gamma (t) - P_\Delta (t) \label{q25}
\end{eqnarray}
\begin{eqnarray}
P_\Lambda &=& - c^2 \rho_\Lambda\label{q26}\\
P_\Gamma &=& c^2\rho_\Gamma /3  =(a/3) T^4(t) \label{q27}\\
P_\Delta (t) &=& c^2\rho _\Delta (t)\omega_\Delta (t)\label{q28}\\
P_\Lambda  &=& P_\Gamma (t) - P_\Delta (t) \label{q29}\\
\rho_\Lambda  &=& \rho_\Delta (t)\omega_\Delta (t)-\rho_\Gamma (t)\omega _\Gamma  \label{q30}
\end{eqnarray}
Already, it is clear that equation (\ref{q30}) defuses the numerical mismatch problem shown at equation (\ref{q0}) because it tells us that Einstein's dark energy density, $\rho_\Lambda$, is not just zero point energy which is believed can assume large values but rather it is the {\it difference\/} of two quantities that even if individually large can give a much smaller value. It is convenient to use Einstein's dark energy density $\rho_\Lambda$ from now on in the discussion. However as explained in $A$, $B$, and $C$, I believe that the quantity $\rho^\dagger _\Lambda =2\rho_\Lambda$ is physically more important.  Thus the main problem of the conflict between the numerical values found for zero point energy from quantum mechanics and dark energy from cosmology does not appear in this cosmological model. Rather, it is shown that dark energy involves the difference of zero point pressures as might be calculated from quantum theory. However, let us explore the relations  further before considering what effect, if any, zero point energy has on the $CMB$ temperature.

Consider replacing the densities with their zero point versions, (\ref{q9}) and (\ref{q19}) in the equation (\ref{q30}). This gives
\begin{eqnarray}
\rho_\Lambda  &=& \rho _{\Delta, \nu _c}  (t)\omega_\Delta (t)-\rho _{\Gamma, \nu _c} (t)\omega _\Gamma.\label{q31}
\end{eqnarray}
The terms can then be expanded into their zero point and nonzero point parts. We obtain,
\begin{eqnarray}
\rho_\Lambda  &=& ( \rho _{\Delta Z, \nu _c}  (t)+ \rho _{\Delta B, \nu _c}  (t))\omega_\Delta (t) -( \rho _{\Gamma Z, \nu _c} (t)+ \rho _{\Gamma B, \nu _c} (t) ) \omega _\Gamma.\label{q32}
\end{eqnarray}
Thus

\begin{eqnarray}
\rho_\Lambda  &=&\ \  (\rho _{\Delta Z, \nu _c}  (t) \omega_\Delta (t))- \rho _{\Gamma Z, \nu _c} (t) \omega _\Gamma)\nonumber\\
&\ & +(\rho _{\Delta B, \nu _c}  (t) \omega_\Delta (t) -\rho _{\Gamma B, \nu _c} (t) \omega _\Gamma ) \label{q33}
\end{eqnarray}
and the quantity $\rho_\Lambda$, known as Einstein  {\it dark energy density\/}, from this cosmological theory is shown to have two parts $\rho_{\Lambda,Z}$ and $ \rho_{\Lambda,B}  $, say, given by 
\begin{eqnarray}
\rho_{\Lambda,Z}&=& \rho _{\Delta Z, \nu _c}  (t) \omega_\Delta (t)-\rho _{\Gamma Z, \nu _c} (t) \omega _\Gamma \label{q34}\\
\rho_{\Lambda,B}  & = &\rho _{\Delta B, \nu _c}  (t) \omega_\Delta (t) - \rho _{\Gamma B, \nu _c} (t) \omega _\Gamma. \label{q35}\\
\rho_\Lambda  &=&\ \rho_{\Lambda,Z} + \rho_{\Lambda,B}\label{q35.1}
\end{eqnarray}
The first part determined by the zero point energy parts of the separate $\Gamma$ and $\Delta$ fields. The second part determined by the usually finite parts of the same two fields. Both of these parts can be finite. The first because it is the difference of two possible large quantities. The results (\ref{q34}) to (\ref{q35.1}) achieve the main objective of this paper giving the definitive result that the dark energy density can be small whilst the zero point contributions can be large. This definitive result is achieved against the difficulty of making reliable calculations of the zero point energy terms and is a direct result of how the $CMB$ pressure arises in this cosmological model. However, in spite of the difficulties I believe that this analysis can be pushed further to give more conclusions that are perhaps not so definitive as follows.  

Equation (\ref{q35.1}) has the direct interpretation that adding into the model the zero point energy versions of the densities results in Einstein's dark energy density, $\rho_\Lambda$, itself acquiring the zero point energy, $\rho_{\Lambda,Z}$. This is a consequence that possibly could have been anticipated. However, it raises the question {\it what is actually measured by the astronomers'  $\Omega _\Lambda$ term?\/} is it just the finite  part $\rho_{\Lambda,B}$ or is it the full $\rho_\Lambda$. It seems to me that this is not obvious but I expect the astronomers will claim it is just the finite part which is usually conceived of as having measurement significance. If that is true and all the measurements involved in equation (\ref{q35.1}) are finite part measurements, then taking the zero point part to be  zero would ensure the consistency of the original equation before adding in the zero point contributions to the theory.

The structure unearthed here does suggest how nature uses the zero point energy of combined fields to embed them together to form a single entity. In this case, the $\Gamma$ and $\Delta$ fields together need to have a common energy reference level and as we have seen before introducing the zero point idea into this work simply subtracting one from the other seemed to be working very well in developing this theory and, in fact, giving the measured result at equation (\ref{q36} ). This suggests that with the addition of the zero point terms equation (\ref{q37}) can be regarded as an auxiliary equation that ensures that equation (\ref{q36}) does represent the measurement as it is usually conceived. 
\begin{eqnarray}
\rho_\Lambda &=& \rho _{\Lambda,B} =\rho _{\Delta B, \nu _c}  (t) \omega_\Delta (t) -\rho _{\Gamma B, \nu _c} (t) \omega _\Gamma \label{q36}\\
\rho_{\Lambda,Z} &=& (P _{\Delta Z, \nu _c}(t) - P_{\Gamma Z, \nu _c} (t) )/c^2  \label{q37}\\
&=& 0\label{q38}
\end{eqnarray}
for Einstein's dark energy. Equations (\ref{q37}) and (\ref{q38}) could lead to the speculation that when zero point energy fields are combined their zero point {\it pressures\/} assume a mechanical equilibrium configuration implied by equation (\ref{q38}). It should also be noted that equation (\ref{q38}) can also be expressed as
\begin{eqnarray}
0&=& \rho _{\Gamma Z, \nu _c} (t) \omega _\Gamma - \rho _{\Delta Z, \nu _c}  (t) \omega_\Delta (t) \label{q39}\\
\frac{\rho _{\Gamma Z, \nu _c} (t)}{ \rho _{\Delta Z, \nu _c}  (t)}  &=&\frac{ \omega _\Delta }{ \omega_\Gamma (t)}.\label{q40}
\end{eqnarray}
From paper $C$, we know that
\begin{eqnarray}
\omega_\Gamma &=& 1/3\label{q41}\\
\omega_\Delta (t) &=& \left(\frac{M_\Gamma }{3 M_U} +\frac{3 (c/R_\Lambda)^2 \rho^{-1}(t)}{8 \pi G} \right)/(1 -M_\Gamma/M_U).\label{q42}
\end{eqnarray}
Thus to get the very simple structure described by such an assumed zero point pressure equilibrium shown at equation (\ref{q36}) the densities would have to be in the ratio described by equations (\ref{q40}), (\ref{q41}), and (\ref{q42}). There is another and equivalent way to ensure that equation (\ref{q39}) holds and that is to use equations (\ref{q39}), (\ref{q41}) and (\ref{q42}) to {\it define\/} the zero point energy of the none $CMB$ part of the cosmos, $\rho _{\Delta Z, \nu _c} (t)$, in terms of the zero point energy of the $CMB$ part, $\rho _{\Gamma Z, \nu _c} (t)$. This would be a very reasonable course of action in view of the extreme difficulties that would be involved in any direct attempt to calculate $\rho _{\Delta Z, \nu _c} (t)$.
\section{Zero Point Energy and Thermodynamics}
\setcounter{equation}{0}
\label{sec-zpet}
The formula relating temperature and time (\ref{E1.200}) is different from those which are used in the standard model. The formula here arises in this model in a very natural way. Again from paper $C$, other thermodynamics quantities for the {\it CMB\/}, the free energy $F(T,V)$  as a function of temperature and volume, entropy $S_\Gamma$, pressure $P_\Gamma$ and energy $E_\Gamma$ also arise naturally in their usual forms and as functions of time,
\begin{eqnarray}
F(T,V) &=& - (a/3) V_U(t)T^4(t)\label{E1.210}\\
S_\Gamma (t) &=& \left (-\frac{\partial F}{\partial T}\right)_V= (4 a/3) V_U(t)T^3(t)\label{E1.220}\\
P_\Gamma (t) &=& \left (-\frac{\partial F}{\partial V}\right)_T= (a/3) T^4(t) \label{E1.230}\\
E_\Gamma  &=&  a V_U(t)T^4(t).\label{E1.240}
\end{eqnarray}
A very clear and accurate description of the thermodynamics of blackbody radiation can be found in {\it F. Mandl's\/} book \cite{49:man} page 260.
Taking the cosmic microwave background radiation to conform to the usual blackbody radiation description, the {\it mass\/} density function for the {\it CMB\/} will have the form
\begin{eqnarray}
\rho_\Gamma (t) &=& a T^4(t)/c^2\label{E1.40}\\
a&=& \pi ^2 k^4/(15 \hbar ^3 c^3)\label{E1.50}\\
 &=& 4\sigma /c,\label{E1.60}
\end{eqnarray}
where $\sigma$ is the Stephan-Boltzmann constant,
\begin{eqnarray}
\sigma  &=& \pi ^2 k^4/(60 \hbar^3 c^2)\label{E1.70}\\
&=& 5.670400\times 10^{-8}\quad W\ m^{-2}\ K^{-4}.\label{E1.80}
\end{eqnarray}
The total mass associated with the  {\it CMB\/} is given by (\ref{E1.90}). That is 
\begin{eqnarray}
M_\Gamma &=&  \rho_\Gamma (t) V_U(t) \label{E1.90}\\
&=& (a T^4(t)/c^2) 4 \pi r^3(t)/3\label{E1.100}\\
&=& (8\pi a T^4(t)/(3 c^4 )) (R_\Lambda )^{2}  M_U G\sinh^{2}(\theta_\pm (t)). \label{E1.121}
\end{eqnarray}
It follows from (\ref{E1.121}) that the temperature, T(t),  as a function of epoch can be expressed as,
\begin{eqnarray}
(8\pi a T^4(t) / (3 c^4 ) )  &=& M_\Gamma /((R_\Lambda )^{2}  M_U G\sinh^{2}(\theta_\pm (t)))\label{E1.130}\\
T(t) &=&\pm((M_\Gamma 3c^2/(4 \pi a))(r(t)^3))^{1/4}. \label{E1.160}
\end{eqnarray}
The well established fact that at the present time, $t^\dagger$, the temperature of the $CMB$ is
\begin{eqnarray}
T(t^\dagger) = T^\dagger =2.725\  K\label{E1.170}
\end{eqnarray}
can be used to simplify the formula (\ref{E1.160}) because  with (\ref{E1.170})
it implies that 
\begin{eqnarray}
0<T(t^\dagger)&=&+(M_\Gamma 3 c^4 /(8\pi a (R_\Lambda )^2  M_U G\sinh^2(\theta_\pm (t^\dagger))))^{1/4}\label{E1.180}
\end{eqnarray}
or
\begin{eqnarray}
T(t)/T(t^\dagger)&=&\pm (\sinh^2(\theta_\pm (t^\dagger)))^{1/4}/(\sinh^2(\theta_\pm (t)))^{1/4}\label{E1.190}\\
T(t) &=&\pm T^\dagger \left(\left(\frac{\sinh(\theta_\pm (t^\dagger))}{ \sinh(\theta_\pm (t)))}\right)^2\right)^{1/4}.\label{E1.200}
\end{eqnarray}
The result of using the  zero point energies for the density quantities on the temperature, $T$, of the $CMB$ given by equation (\ref{E1.160}), in the case of this model can easily be found. Inspection of equation  (\ref{E1.130}) reveals that the only change could come in through the terms $M_\Gamma$ and $M_U$ which would switch to $M_{\Gamma, n_c}$ and $M_{U,n_c}$. However, when we use the measured value $T^\dagger$ for the temperature of the $CMB$ at equation (\ref{E1.170}) we see that these constants cancel out completely to give the final formula for $T(t)$ at equation (\ref{E1.200}). Thus, conclusively, the $CMB$ temperature does not involve the zero point quantities and this implies that the temperature measurements imply that they are effectively arising from the thermodynamic energy field $E_\Gamma$ just as in (\ref{E1.240}) or equivalently from its density $E_\Gamma /(V_U (t))$. This is an important result as it appears to set the tone for measurement interpretation generally as far as this model is concerned. It seems that measurements do not involve the zero point contributions. Returning to the thermodynamic equations (\ref{E1.210}) to (\ref{E1.240}) we can consider the possibility that the zero point contributions may contribute a term $Z(t)$ that does not explicitly depend on $T$ or $V_U$  to the free energy term with consequences for the other terms so that equation (\ref{q43}) holds with the following results.
\begin{eqnarray}
F(T,V) &=& - (a/3) V_U(t)T^4(t) + Z(t)V_U(t) \label{q43} \\
S_\Gamma (t) &=& \left (-\frac{\partial F}{\partial T}\right)_V= (4 a/3) V_U(t)T^3(t),\label{q44}\\
P_\Gamma (t) &=& \left (-\frac{\partial F}{\partial V}\right)_T= (a/3) T^4(t) -Z(t),\label{q45}\\
E_\Gamma  &=&  a V_U(t)T^4(t)- 3Z(t)V_U(t).\label{q46}
\end{eqnarray}
Thus we conclude that adding in the zero point energies does {\it not\/} change the entropy but does change the pressure and the energy. The energy change is really where we started by introducing the zero point energy terms into the scheme. However, it might be thought that the formula for temperature obtained from equation (\ref{E1.100}) is now not consistent with the structure. However, this is no problem as clearly the temperature should be obtained from the temperature dependent part of $E_\Gamma$ and that indeed is what is used in that equation and so it should be retained in that form. The pressure does have an extra zero point part which again is not unexpected and it may mean that numerical theoretical values for pressures should be adjusted. However, if the proposition at equation (\ref{q38}) does turn out to be correct, zero point {\it pressures\/}  for the total system cancel to maintain the two, $\Delta$ and $\Gamma$, subsystems in mechanical equilibrium.
\section{Conclusions}
The main result derived in this paper, using the dust universe cosmology model with components that have zero point energies, is that Einstein's dark energy density $\rho_\Lambda$ arises itself with a zero point energy contribution $\rho_{\Lambda,Z}$ as the difference of two zero point energy contributions (\ref{q47}) or (\ref{q48}),
\begin{eqnarray}
\rho_{\Lambda,Z}&=& \rho _{\Delta Z, \nu _c}  (t) \omega_\Delta (t)-\rho _{\Gamma Z, \nu _c} (t) \omega _\Gamma \label{q47}\\
\rho_{\Lambda,Z} &=& (P _{\Delta Z, \nu _c}(t) -P_{\Gamma Z, \nu _c} (t) )/c^2  \label{q48}\\
\rho_{\Lambda,B}  & = & \rho _{\Delta B, \nu _c}  (t) \omega_\Delta (t) -\rho _{\Gamma B, \nu _c} (t) \omega _\Gamma \label{q49}\\
\rho_\Lambda  &=&\ \rho_{\Lambda,Z} + \rho_{\Lambda,B}.\label{q50}\\
c^2 \rho_\Lambda &=& - P_\Lambda \approx (P_{\Delta B, \nu _c} (t)-P _{\Gamma B, \nu _c}(t)) \label{q51}
\end{eqnarray} Thus although the zero point energies could, as indicated from quantum theory, turn out to be very large $\rho_{\Lambda,Z}$ could turn out to be small, very small or indeed zero. Thus the work here resolves the large number mismatch mystery. It is further suggested that the extra zero point term $\rho_{\Lambda,Z}$ that adds on to the usual finite part $\rho_{\Lambda,B}$ denoted by the $B$ subscript can be taken to be zero as a form of auxiliary condition that defines the energy zero and the  usual perception of the meaning of measurements. However, this suggestion cannot at this time be confirmed because reliable calculations of zero point energies involved cannot be made. However, the zeroness of $\rho_{\Lambda,Z}$ would imply that the zero point contributions have to be in mechanical pressure equilibrium (\ref{q48}) to ensure the stability of their vacuum togetherness.
It should be noted that equation (\ref{q49}) suggests that $\rho_\Lambda$ is just a {\it weighted\/} relative density of the $\Delta $ field with respect to the $\Gamma$ field and that equation (\ref{q51}) implies that $P_\Lambda$ is the relative pressure of the of the $\Gamma$ and $\Delta$ fields.

Expressed otherwise the remarks in the previous paragraph about the approximate equation (\ref{q51}) that could turn out to be exact suggests that there is a simple physical interpretation for the dark energy quantities $\rho _\Lambda$ and $P _\Lambda$ and the part they play in relation to the universe's structure and evolution in terms of more familiar fields.  The constant valued everywhere and at all time quantity $P_\Lambda$ can be regarded as a measure of the resultant of a {\it pressure\/} polarisation into a positively valued $\Delta$ field material pole and a negatively valued anti-pole formed from $\Gamma$ field material. A similar remark applies to the $\rho_\Lambda$ quantity by equation (\ref{q49}) which just reflects the same pressure polarisation. These polar quantities occupy the same interior to the universe space and are not spatially separated. These remarks apply inside the universe. Outside the universe, the $\Lambda$ quantities  exist in their un-polarized constant condition. Thus the whole expansion process of the universe can be thought of as a spherically spreading {\it polarisation\/} of a hyper-vacuum by activated dark energy converting into $\Delta$ and $\Gamma$ polar components. I emphasise that the {\it pressure\/} polarisation concept does not imply physical mass density polarisation. However, by multiplying equation (\ref{q51}) through by $G$ the polarisation process can more appropriately be seen to be gravitational polarisation,
\begin{eqnarray}
G\rho_\Lambda = G_{-}\Gamma_B  + G_{+} \Delta _B, \label{q52}
\end{eqnarray}
where $ G_{-}=-G$ and $ G_{+} =G$.
 \section{Appendix 1 Standard Cosmology Model Structure}
\large
\centerline{\Large {\bf Alternative to the Epoch Time Solution Patching }}
\centerline{\Large {\bf Structure of the Cosmological Standard Model }}
\centerline{\Large {\bf in the Friedman Dust Universe}}
\centerline{\Large {\bf with Einstein's Lambda}}
\vskip0.5cm
\centerline{April 14, 2010}
\vskip 0.75cm  
\section{Appendix 1 Abstract}
The need for the cosmological constant, Lambda, in Einstein's field equations to be an absolute mathematical constant over all the time that they are used to describe some astrophysics process is demonstrated. Only if that condition holds will the conservation laws of mass and momentum hold as in classical physics. The Friedman equations that can be deduced rigorously from general relativity are consequently equally restricted to a constant valued Lambda and for the same reasons. However, the standard cosmological model, is not constructed from one solution of the Friedman equations but rather from at least three different but rigorous solutions patched together at times where they are physically thought to join. This is because the known solutions are thought to represent different conditions of mass movement, highly erratic or thermal at time near the big bang or more particle like and organised into systems at time near now, just to mention two types of activity when there obviously could be a continuous range of activities of mass types. Clearly this idea of how things have evolved after the big bang is very plausible, if the big bang idea is accepted as fact. The apparent need to patch solutions together over time creates great mathematical difficulties for cosmology because the three functions selected have to join smoothly which is the same as saying that they have to be differentiable not once but twice if accelerations are taken into account as they must be if the Friedman equations are to hold through the join. It is not clear whether or not this patching process can be rigorously achieved. However it is clear that the big bang concept does violate Einstein's field equations at $t=0$ because this concept implies that mass and momentum comes from nowhere. It is shown that all of these problems can be removed by introducing a continuously variable over time structure into the definition of temperature for the dust universe model.
This only affects the value of the temperature that is associated with a given time and make no difference to the validity of the dust universe model with regard to it being a rigorous solution to the Einstein Field equations for all time from minus infinity to plus infinity. 
\vskip 0.2cm 
\centerline{Keywords: Cosmology, Dust Universe, Dark Energy, Epoch}
\centerline{Time Solutions Matching, Friedman Equations, Standard Model}
\centerline{Einstein's field Equations, Lambda}
\vskip 0.3cm

\section{Appendix 1 Introduction}
\setcounter{equation}{0}
\label{sec-introefe}
In the following pages, I shall introduce a simple extension to the dust universe model that greatly enlarges its ability to describe astrophysical phenomena that are conceived as depending strongly on the cosmological temperature. This is particularly relevant to those processes that involve transitions from heat dominated disordered conditions at early epoch time to the cooler mass particle ordered conditions of the present time. I shall first consider the important contribution of the cosmological constant in terms of its contribution to the conservation of mass and momentum as described by the Einstein field equations.  

In $1917$, Einstein introduced his modified  tensor field  equations, (\ref{1}),(\ref{3}) and (\ref{4}), with the addition of the so called {\it Lambda\/}, $\Lambda$, term, 
\begin{eqnarray}
R_{\mu\nu} -\frac{1}{2}g_{\mu\nu}R +g_{\mu\nu}\Lambda &=& -\kappa T_{\mu\nu}.\label{1}\\
G_{\mu\nu }&=&R_{\mu\nu} -\frac{1}{2}g_{\mu\nu}R\label{2}\\
G^\mu_{\ \nu;\mu}&=&0\label{3}\\
T^\mu_{\ \nu;\mu}&=&0.\label{4}
\end{eqnarray}
Equation (\ref{2}) is the definition of the Einstein tensor, $G_{\mu\nu}$, to be used in the line below it. Equations (\ref{3}) and (\ref{4}) are covariant derivatives of $G_{\mu\nu}$ and the stress energy momentum tensor, $T_{\mu\nu}$, respectively. They are both first order tensor, or equivalently vector equations, because one index of a second order tensor has been contracted out through the differentiation process. They both represent conservation of energy and momentum. The zero character of the covariant derivative of $ G^\mu_{\ \nu;\mu}$ is an inevitable consequence of the geometrical structure represented by $ G_{\mu\nu }$ and the way it is defined. The zero character $T^\mu_{\ \nu;\mu}$ is an inevitable consequence of the physical meaning of $T_{\mu\nu}$ and the way it is represented. Thus it follows that with the exception of the $\Lambda$  term all the terms in that equation, (\ref{1}), satisfy the laws of energy and momentum  conservation of classical physics. That is to say Einstein's {\it original unmodified\/} equation conforms to classical energy momentum conservation. However even this conclusion is not completely true for the following reason. The time variable that emerged with Einstein's original equations has an unusual property in classical physics experience. The Einstein field equations have a built in singularity at cosmological time zero which of course, we all know about as the big bang origin of the universe. However, proponents of the big bang theory accommodate the massive violation of mass and momentum conservation at the instant $t=0$ by the caveat that everything began at that instant or just after  so that the field equations and the covariant conservation equations can be disregarded at that instant. So ignoring this complication temporarily, at time other that time zero we now need to consider what the situation is when the $\Lambda$ term is left in the tensor field equation (\ref{1}) as in the modified form.
For the Einstein field equations with the lambda term included not to be in conflict with the energy momentum conservation laws of classical physics it is necessary that the additional term $g_{\mu\nu}\Lambda$ also satisfies the covariant conservation equation,
\begin{eqnarray}
(g^\mu_{\ \nu}\Lambda)_{;\mu}&=&0.\label{5}
\end{eqnarray}
That is
\begin{eqnarray}
(g^\mu_{\ \nu})_{;\mu}\Lambda + g^\mu_{\ \nu} ( \Lambda)_{;\mu}&=&0\label{6}\\
g^\mu_{\ \nu} ( \Lambda)_{;\mu}&=&0.\label{7}
\end{eqnarray}
The last equation following because the covariant derivatives of the $ g_{\mu\nu}$ are zero. The Cosmological quantity $\Lambda$
is never taken to depend on the space variables so that the only derivative left to consider is the time variable, $x_4$ differentiation part in (\ref{6}), 
\begin{eqnarray}
g^4_{\ \nu} ( \Lambda)_{;4}&=& g^4_{\ \nu} ( \Lambda)_{,4}=0,\label{8}
\end{eqnarray}
because the covariant derivative of a scalar is equal to the normal derivative.
There will certainly be at least one of the $ g^4_{\ \nu} $ elements that is not zero. It therefore follows that the Lambda included version of Einstein's field will only conform with the conservation laws of classical physics provided that the $x^4$ or time derivative of $\Lambda$ satisfies
\begin{eqnarray}
\partial \Lambda/\partial t=0.\label{9x}
\end{eqnarray}
That is to say that the cosmological quantity $\Lambda$ has to be an absolute constant. If a space or time variable $\Lambda$ is used in Einstein's field equations their physical mathematical validity is totally compromised. Lambda is rightly referred to as {\it The Cosmological Constant\/}. In the next section, I examine how the covariant conservations laws of Einstein's general relativity are reflected in the structure of the cosmological standard model.
\section{Standard Cosmology Model}
That something very strange happens at cosmological time zero seems to be an unavoidable consequence of Einstein's field equations for general relativity. This feature becomes very noticeable in the consequent Friedman equations and their solutions derived from relativity and used for constructing cosmological models. However, the formulation of these equations has no restriction on the possible $-\infty\rightarrow +\infty$ time range. We have all heard about this initial extraordinary event called either the big bang or the initial singularity. The ideas that go along with the big bang concept are expansion of the universe from a zero or very small volume at the same time as a very large quantity of mass possible the whole mass of the universe, $M_U$ is generated within this volume, a very violent {\it explosion\/} from a very small volume so that the pressures and temperatures involved must have been enormous, rather like an incredibly large atomic explosion. This image leads to the idea that the initial stages of the universe must have been dominated by radiation with an excess of photon like activity. Conceivable then as expansion proceeded cooling and pressure reductions occurred leading to condensation of matter into particulate forms with very much less kinetic activity. It seems to me that these ideas but with very little supporting evidence from actual observation are the reason that cosmological models starting with the big bang are taken to have three main time phases. They start with inflation, an attempted explanation of the start from nothing, followed by a radiation dominated phase which is followed by a matter dominated phase. Now days the model has to take account of a recently observed accelerated radial expansion of the universe. At some $t_c$ in the evolutional time sequence above an initial deceleration condition due to gravitation attraction is assumed to change into an accelerating condition due to the universe's dark energy content increasing with time. This type of prescription for the time evolution of the universe has for years presented cosmologist with a dilemma because although a number of solutions to the Friedman equations have been found they all seemed to have different characteristics with regard to whether they represented radiation or matter evolving with time.  Thus in constructing the standard model for cosmology it seemed imperative that different solution would have to be time wise patched together if the history envisaged above were to be mathematically represented in the standard model. This problem is now complicated by the problem of incorporating the so described mysterious dark energy. So it was that mathematically solutions of the Einstein field equations were time wise patched together to give a model that it is claimed represents the actual time evolution of the universe from time about zero to time in the unlimited distant future, $t=+\infty$. However, there is a down side to this approach which is first of all the big bang concept and the inflationary beginning which obviously breaks Einstein's conservation rules at time zero or thereabouts by the generation of mass from nothing and this is confounded by the inflation section which lasts for a finite time, supposedly starting before time zero and involving a massive value for the cosmological constant which clearly cannot be matched with the small values following at later times and so must violate the time constancy of $\Lambda$ condition obtained at (\ref{9x}). The later time sections also have to be matched and should involve $\Lambda$ not changing, smooth connections between other parameters and their time derivatives should happen at these times.  I do not wish to be dogmatic but it seems to me that this smooth connection scenario is not mathematically demonstrably achieved. Thus ignoring the problem with the inflation section on the grounds that it is inevitable if the big bang concept is true, big problems at later times are still present in the standard model. I shall show in the next section that there is a way of avoiding the uncertainties of the standard model by not using time patching at all. 
\section{Dust Universe Model}
\setcounter{equation}{0}
\label{sec-intro}
In an early stage of introducing the dust universe cosmological model, I imposed a restriction on the relation between two types of mass into which the universe can be divided, the thermal mass of the cosmic micro-wave background and the rest which I denoted by $M_\Gamma$ and $M_\Delta$ respectively, so that the total non-dark energy mass $M_U$ could be represented as
\begin{eqnarray}
M_U =  M_\Delta + M_\Gamma.\label{1a}
\end{eqnarray}
In this model I define $M_U$, the total mass within the spherical boundary of the universe not including any dark energy mass to be taken as being conserved that is of retaining the constant numerical value throughout the time history of the evolution of the universe for this model. However additionally,    
I imposed the working condition that both of the masses, $M_\Delta$ and $ M_\Gamma $ are to be separately conserved that is their numerical values are to remain constant throughout the time history of the evolution of the universe for this model. I called this restriction on the two mass components the {\it strong\/} assumption and in fact I introduced it in the mistaken belief that it made progress with the theory possible which would be otherwise very difficult. The physical consequence of this assumption was that the dust universe model had a history in which total amount of cosmic background mass $M_\Gamma$ has to remain constant even though its local density can change with time. Clearly this is closely related to the problem of the necessity of time wise patching different radiation or matter dominating solution to get a complete model. Returning to this issue a few years later, I have found that this restriction can easily be removed making no difference to the basic form of the physics or mathematics of the structure. The need to replace this restriction was pointed out to me by Professor C. W. Kilmister some years ago. There is a big bonus earned in making this change. I shall show in the next section that the original model can easily be replaced by model with one rigorous solution to the field equations holding over all time, $-\infty\rightarrow +\infty$ but which can change its matter character type  as it evolves with time. This is achieved by  making a minor addition to the original model related to the definition of temperature. This addition then becomes an input function of a time parameter that users of the theory can chose to fit any theoretical or measured continuous time sequence of mass character that they decide best fits an evolving cosmology universe model.
Let us consider the definitions of mass density and temperature from the dust universe model (\cite{45:gil})
\begin{eqnarray}
\rho (t)&=& (3/(8\pi G))(c/R_\Lambda)^2(\sinh^{-2}(3ct/(2R_\Lambda))\label{2a}\\
T(t)&=&\pm (M_\Gamma 3 c^4 /(8\pi a (R_\Lambda )^2  M_U
G\sinh^2(3ct/(2R_\Lambda))))^{1/4}\label{3a}\\
T^4 (t)&=&\pm (M_\Gamma  c^2 \rho (t)/(a M_U
)) \label{4a}\\
\end{eqnarray}
The third equation above arises from the first two and gives the temperature at time $t$ in terms of the density at time $t$. The total mass of the cosmic microwave background radiation in the universe, $M_\Gamma$, in this formula is taken to be a constant in that theory. Thus we were able to write the ratio of the temperature at two different times $t_1$ and $t_2$ as equal to the ratio of the fourth roots of the density at the same two times as,
\begin{eqnarray}
\left(\frac{T(t_1)}{T(t_2)}\right)^4=\left(\frac{\rho (t_1)}{\rho (t_2)}\right). \label{5a}
\end{eqnarray}
On the other hand, if $M_\Gamma$ is some arbitrary function of time, $M_\Gamma(t)$, say, then we have to replace the formula (\ref{5a}) with(\ref{6a}), 
\begin{eqnarray}
\left(\frac{T(t_1)}{T(t_2)}\right)^4= \left(\frac{ M_\Gamma (t_1) \rho (t_1)}{\ M_\Gamma (t_2) \rho (t_2)}\right)&=& \Gamma (t_1,t_2) \left(\frac{ \rho (t_1)}{ \rho (t_2)}\right) \label{6a}\\
\Gamma (t_1,t_2)&=& \frac{ M_\Gamma (t_1)}{ M_\Gamma (t_2) } \label{7a}\\
\Gamma (t_1,t_2)&=& \Gamma^{-1} (t_2,t_1).\label{8a}\\
\Gamma (t_2,t_1) \left(\frac{T(t_1)}{T(t_2)}\right)^4&=& \left(\frac{ \rho (t_1)}{ \rho (t_2)}\right). \label{9a}
\end{eqnarray}
From (\ref{2a}) and (\ref{3a}) it can be seen that making $M_\Gamma$ time dependent can have no effect on $\rho (t)$ as $M_\Gamma$ only occurs in the temperature, (\ref{3a}). Also, the temperature is not used to define any other quantities in the theory so that the transition from the old theory to its modification can be read off from (\ref{9a}) as replace all occurrences in the original theory of the temperature times ratio in (\ref{6a}) with the modified ratio form as indicated as below
\begin{eqnarray}
\left(\frac{T(t_1)}{T(t_2)}\right)^4\rightarrow\Gamma (t_2,t_1) \left(\frac{T(t_1)}{T(t_2)}\right)^4.\label{10a}
\end{eqnarray}
The only restriction on choosing the function of time $M_\Gamma (t)$ for making this transition is that it will have to conform to (\ref{1a}) as below
\begin{eqnarray}
M_U =  M_\Delta (t) + M_\Gamma (t).\label{11a}
\end{eqnarray}
That is to say that because $M_U$ is an absolute constant $M_\Delta$ will have to depend on $t$ and
\begin{eqnarray}
M_\Gamma (t)&\le& M_U \ \forall t\label{12a}\\
M_\Delta (t)&=& M_U - M_\Gamma (t)\ \forall t.\label{13a}
\end{eqnarray}
Otherwise the mass type conditioning function with time, $M_\Gamma (t)$,
is {\it arbitrary\/}.
This structure with its infinite time range continuous adaptability is obviously a great advance over the standard model time patching scheme, only applicable in about three finite time ranges, with very difficult mathematical problems associated with joining time sectors. 
\section{Appendix 1 Conclusions}
\setcounter{equation}{0}
\label{sec-con3}
The mathematical uncertainties of the joining of differently physically characterised time range solutions of the Friedman equations used in the standard cosmological model can be bypassed using a single solution suitable adapted with regard to its definition of temperature. The single solution used to make the case for this change is the {\it dust universe model with Einstein's Lambda\/} which is a rigorous solution to Einstein's Field equations over all time, $-t_\infty\rightarrow +t_\infty$. The mass and momentum conservations laws hold at all times for this model so that additionally to any phase matching avoidance at time other than zero the conceptual and mathematical problems of the inflation phase of the standard model are also avoided.
\newpage 
\section{Appendix 2: Dark Matter}
\vskip 0.75cm  
\large
\centerline{\Large {\bf The Dark Matter Problem}}
\centerline{\Large {\bf General Relativistic Galactic Rotation Curves}}
\centerline{\Large {\bf in a Friedman Dust Universe}}
\centerline{\Large {\bf with Einstein's Lambda}}
\vskip0.5cm
\centerline{June 3 2010}
\vskip 0.75cm  
\section{Appendix 2 Abstract}
In this paper, the general relativistic replacement for the Newtonian inverse square law of gravitation is obtained from the Friedman Cosmology equations. This version of the inverse square law is shown to contain information about the amount of dark energy mass contained in a specific region through a mass term $M_\Lambda^-$ dependent on Einstein,s Lambda and, importantly for this paper, it also contains information about the amount of dark matter mass in the same region through a term $M_P^+$. This work derives from the Dust Universe Model which gives a complete cosmological description of the movement and evolution of the astrophysical space substratum which as usual is represented by a spatially uniform or constant mass density distribution at zero pressure. Thus definite spatial regions of the substratum can only be regarded as holding regions for un clumped mass, as primitive galaxies might be described. Consequently, to describe actual galaxies that have  condensed from such a region, the more general solution of Einstien's Field eqtions involving the pressure term is needed to explain clumping and the resultant galactic form. The general relativist version of the inverse square law is written in a form applicable to the case of bound circular orbiting about a spherically symmetric central gravitational spatially distributed source force. Thus the behaviour of masses cycling within or outside the source region can be analysed. The formula for the galactic rotation curves for stars rotating within or outside the source region is obtained.  A very simple galactic model is used consisting of just two components, the halo and the bulge with all visible orbiting stars,  The conclusion is that the pressure term from general relativity and in the consequent Friedman equations is adequate to explain the constancy of the function of rotational velocity as a function of orbital distance from the centre of gravity starting at the massive core of the galaxy. A simple and parameter adaptable computer program using Mathematica has been constructed to display diagrams of galactic rotation curves. This program is available for downloading.     
\vskip 0.2cm 
\centerline{Keywords: Cosmology, Dust Universe, Dark Energy, Dark matter}
\centerline{Galactic Rotation Curves, Friedman Equations}
\centerline{General Relativity, Pressure, Inverse Square Law}
\vskip 0.3cm

\centerline{PACS Nos.: 98.80.-k, 98.80.Es, 98.80.Jk, 98.80.Qc}
\section{Appendix 2 Introduction}
\setcounter{equation}{0}
\label{sec-introefe2}
In the following pages, I shall introduce a general relativistic substructure into the dust universe model that can be used to describe galactic rotation phenomena and show how these relatively stable massive systems, having condensed from the local substratum expansion, can exhibit the constant with respect to radial distance rotation curves that have recently been observed. The dust universe model is left completely intact by this addition but local redistributions of the uniform mass within the expanding substratum mass platforms of that model can now rearrange and, in particular, can clump to form mass distributions separated from neighbouring platforms. The total mass involved in any such redistribution will remain constant. This is achieved by using a general relativistic generalisation of Newton's inverse square law of gravitation at the local galactic level. Thus the {\it dark matter problem\/} can be resolved using standard general relativity theory.
\section{Relativity Generalised Inverse Square Law}
\setcounter{equation}{0}
\label{sec-grgn}
\vskip0.3cm
This generalisation is most simply represented by an equation that can be obtained from the Friedman equation for the acceleration field due to density and is
\begin{eqnarray}
\frac{\ddot r (t) }{ r (t)}  = \frac{\Lambda c^2 }{3} - \frac{4\pi G}{3} \left( \rho (t) + \frac{3 P (t)}{c^2} \right). \label{1c}
\end{eqnarray}
This equation includes a contribution from the Lambda term, $\Lambda c^2 $.
This is a very important equation in relation to the acceleration due to gravity at radius $r$ and how that depends on the mass density term $\rho (t) $. Clearly, the pressure term $3P/c^2$ adds to the mass density to produce an effective or physical  mass density $3P/c^2+\rho $. In the dust universe model, the pressure term is taken to be zero at all times $t$ so that in the dust universe model case the equation above can be written as
\begin{eqnarray}
\frac{\ddot r (t) }{ r (t)}  = \frac{\Lambda c^2 }{3} - \frac{4\pi G}{3} \rho (t) .\label{2c}
\end{eqnarray}
If we define a time dependent mass density, $\rho_P (t)$, that includes the pressure term as 
\begin{eqnarray}
\rho_P(t)=\left( \rho (t) + \frac{3 P (t)}{c^2} \right), \label{3c}
\end{eqnarray}
the original equation (\ref{1c}) can be represented as 
\begin{eqnarray}
\frac{\ddot r (t) }{ r (t)}  = \frac{\Lambda c^2 }{3} - \frac{4\pi G}{3} \rho_P (t)= \frac{\ddot r_P (t) }{ r_P (t)} . \label{4c}
\end{eqnarray}
Thus the alternatively expressed original equation (\ref{4c}) that includes pressure is indistinguishable from the dust universe form of  equation (\ref{2c}) that does not involve pressure, except for the subscript $P$ on the density function. It follows that the solution for $r(t)$ that is involved in the dust universe model is the same as that for the more general non dust models with the plain $\rho (t)$ replaced by $\rho_P(t)$. To avoid confusion when using the more general case, I shall use the subscripted form of the equation given by the second equality in equation (\ref{4c}) above. It is possible to be rather more precise about the meaning of this formula and also show that it is indeed a generalisation of Newton's inverse square law as follows. In its original form as derived from the Friedman equations the non $\Lambda$ part of this formula represents the radial acceleration just outside a sphere of radius $r(t)$ due to the gravity of a spatially uniform distribution of positively gravitating mass centred on a sphere of radius $r(t)$. The $\Lambda$ part of this formula represents the radial acceleration just outside a sphere of radius $r(t)$ due to the gravity of a spatially uniform distribution of density $\rho_\Lambda^\dagger$ of negatively gravitating mass  centred on the same sphere. If we use the volume, $V_P(t)$, of the sphere of radius $r _P (t)$, the formula for $ \rho_\Lambda^\dagger$ and the formulas for total positively gravitating mass $M_P^+$ and total negatively gravitating mass $M_P^-$ within the volume at time $t$ together with the definitions of the gravitational coupling constants for positively $G_+$ and negatively gravitating material $G_-$ respectively given below
\begin{eqnarray}
V_P (t)&=&4 \pi r_P ^3(t)/3\label{5c}\\
\rho_\Lambda^\dagger&=&\frac{ \Lambda c^2}{4\pi G}\label{6c}\\
M_P^+&=& \rho_P (t) V_P (t) \label{7c}\\
M_P^-(t)&=& \rho_\Lambda^\dagger V_P (t) \label{8c}\\
G_+ &=& +G\label{9c}\\
G_- &=& -G\label{10c}
\end{eqnarray}
these definitions can be used to express (\ref{4c}) in the form
\begin{eqnarray}
\ddot r_P (t) = -\frac{G_- M_P^-(t)}{ r^2_P(t) } - \frac{ G_+ M_P^+}{ r^2_P(t) }. \label{11c}
\end{eqnarray}
Expression (\ref{11c}) is the general relativity generalisation for Newton's inverse square law of gravitation that is implied by Einstein's field equations with $\Lambda$. This generalises Newton's inverse square law in three respects. Firstly, the radius variable $r_P (t)$ can depend on time. Secondly a contribution of negatively gravitating material is taken into account through the time dependent mass term $M_P^-(t)$ and thirdly the additional positively gravitational mass due to pressure is taken into account through the non time dependent mass term $ M_P^+$. The derivation above depends on both types of mass density not depending on space variation. However, once this formula is obtained that restriction can be removed because the well known result from Newtonian gravitation theory that says that the mass distribution within the volume with radius $r(t)$ at some fixed time $t$ can be redistributed in any way, with the formula remaining valid, provided its numerical value remains constant and its centre of mass remains at the centre of the sphere.
Reverting back to the original form of the formula (\ref{4c}), we have  
\begin{eqnarray}
\frac{\ddot r (t) }{ r (t)}  = \frac{\Lambda c^2 }{3} - \frac{4\pi G}{3} \left( \rho (t) + \frac{3 P (t)}{c^2} \right) \label{13c}
\end{eqnarray}
and this can be claimed to be the same thing as equation(\ref{11c}), the general relativistic generalisation of Newton's law of gravitation.
This formula can certainly be used in the context of studying well know problems in classical gravitation theory to see what differences the general relativity structure from which it emerged brings to the classical solutions. The objective of this paper is to do just that in the case of the dark matter problem of galactic dynamics. As shown above the formula  (\ref{13c}) can be applied to the case of the gravitational field generated by the fixed amount of mass $ M_P^+$ contained within the time variable volume $V_P (t)$. However, using the purely spherical symmetric case for all functions for simplicity of presentation, it can also be seen to give the general relativistic generalisation of Newton's law that would apply to the more familiar case of the gravitational effect of a fixed amount of mass within a fixed volume when the radial component of velocity $v_1(t)$ is zero and the transverse accelerations $\alpha_2(t)$ is zero, just by dropping the time dependence of $r(t)$  and assuming that $\rho (t)\rightarrow\rho (r,r_i)$ and $P (t)\rightarrow P(r,r_i^\prime)$ and assuming the density and pressure are constant for radii $r$ up to  radii $ r_i $ and $ r_i^\prime $ respectively and zero at greater radii. This last change is certainly a simplifying device to avoid mathematical complications. In effect this is making use of a very simplified model for the mass distribution in a galaxy. One sphere of uniform mass density $ \rho (r,r_i)$, extending from $r=0$ to represent the central bulge and its orbiting stars of radius $r_i$ and a second concentric uniform mass density $\rho (r, r_i^\prime)$  sphere extending from $r=0$ of radius $r_i^\prime $ to represent the halo with $r_i^\prime > r_i$. In more detail these two overlapping densities are defined as,
\begin{eqnarray}
\rho(r,r_i)&=& 0,\ r< 0   \label{13c1}\\
\rho(r,r_i)&=& \rho_{r_i} = a\  constant,\ r\le r_i   \label{13c12}\\
\rho(r,r_i)&=& 0,\ r> r_i   \label{13c13}\\
\rho (r,r_i^\prime ) &=&0,\ r< 0     \label{13c2}\\
\rho (r,r_i^\prime ) &=&\rho_{r_i^\prime}= a\  constant,\ r\le r_i^\prime      \label{13c22}\\
\rho (r,r_i^\prime ) &=&0,\ r> r_i^\prime .      \label{13c23}
\end{eqnarray}
The current thinking on this problem is that most if not all of the missing mass is within the halo. In the case of radial velocity zero, we get    
\begin{eqnarray}
\frac{\ddot r (t) }{ r (t)}\rightarrow -\frac{v^2_2}{r^2} &=& \frac{\Lambda c^2 }{3} - \frac{4\pi G}{3} \left( \rho(r,r_i)  + \frac{3 P(r,r_i^\prime)}{c^2} \right) \label{14c}\\
v_1(t)&=&\dot r(t)=0\Rightarrow r(t)=r=\ a\ constant \label{14c1}\\
v_2(t)&=& r\dot\theta (t)= r\omega\Rightarrow v_2(t)\rightarrow v_2(r) \label{14c2}\\
\alpha _1(t)&=& \ddot r -r\dot\theta ^2(t)= -r\dot\theta ^2(t)\label{15c}\\
\alpha _2(t)&=&r\ddot\theta(t) +2 \dot r = r\ddot\theta(t)  =\frac{\partial(r^2\dot\theta(t))}{r\partial t}=0 \label{15c1}\\
&\Rightarrow& r^2\dot\theta(t)=l=\ a\  constant\label{16c}\\
&\Rightarrow& \dot\theta(t)= \omega =\ a\ constant.\label{17c}
\end{eqnarray}
Writing equation (\ref{14c}) out again in terms of the masses involved inside the spherical volume of radius $r$ we have
\begin{eqnarray}
\frac{v^2_2(r)}{r^2} &=&  \frac{4\pi G}{3} \left( \rho(r,r_i)  + \frac{3 P(r,r_i^\prime)}{c^2} \right)-\frac{\Lambda c^2 }{3}  \label{18c}\\
v^2_2(r) &=&  \frac{G V(r)}{r} \left( \rho(r,r_i)  + \frac{3 P(r,r_i^\prime)}{c^2} \right) -\frac{G V(r) \rho_\Lambda^\dagger}{r}.  \label{19c}
\end{eqnarray}
Let us now define three quantities of mass for the positively gravitating density term $ \rho(r,r_i) $ the pressure term $3 P(r_i,r_i ^\prime)/c^2$ and the dark energy term $\rho_\Lambda^\dagger $ within the spherical volume  $V(r)$ as follows
\begin{eqnarray}
M^+(r, r_i) &=&V(r)  \rho(r,r_i) \label{20c}\\
&=& M^+( r_i, r_i),\  r\ge r_i \label{20c1}\\
M_P(r, r_i^\prime)&=& V(r) \frac{3 P(r,r_i ^\prime)}{c^2} \label{21c}\\
&=& M_P(r_i^\prime, r_i^\prime),\  r\ge r_i^\prime \label{21c1}\\
M_\Lambda(r) &=& V(r) \rho_\Lambda^\dagger .  \label{22c}
\end{eqnarray}
At step (\ref{21c1}) $r$ is outside both distributions of mass.
Using this notation, we can write equation (\ref{19c}) as follows and compare it with the classical velocity circulation equation at the line (\ref{26c}) below, 
\begin{eqnarray}
v^2_2(r) &=&  \frac{G}{r} \left(M^+( r, r_i)  + M_P(r, r_i^\prime)\right) -\frac{G M_\Lambda(r)}{r}   \label{23c}\\
&=&  \frac{G M^*(r)}{r} \label{24c}\\
M^*(r) &=&  M^+( r, r_i)  + M_P(r,r_i^\prime) -M_\Lambda(r)   \label{25c}\\
v^2_2 (r)&=&  \frac{G M(r)}{r}. \label{26c}
\end{eqnarray}
Note that $ M^*(r)$ is only constant for values of $r$ such that $r> r_i^\prime$ when the $M_\Lambda^\dagger$ is excluded.
In the classical case, $M(r)$ means the total amount of mass within a spherical volume of radius $r$ about the radial origin and by construction $ M^*(r)$ means the same thing. The minus sign in front of the $\Lambda$ mass does not mean that this mass is negative, rather it goes with the $G$ at (\ref{24c}) to give the negatively gravitational coupling constant $G_-=-G$ involved with negatively gravitating mass. The expression $M^*(r)$ is a mathematical convenience. The classical version of the formula for boundary velocity in terms of enclosed gravitating mass,  (\ref{26c}), has been used to find the amount of mass $M(r)$ within a spherical region of radius $r$ in terms of its transverse outer edge velocity $v_2(r)$, (\ref{27c}), and also to give the boundary transverse velocity at the edge of an enclosed amount of mass, $M(r)$, (\ref{28c}),
\begin{eqnarray}
M(r)&=&  \frac{v^2_2(r)r}{G}\label{27c}\\
v_2(r) &=&  \left(\frac{G M(r)}{r}\right)^{1/2}.\label{28c}
\end{eqnarray}  These formulae are applicable to planetary systems and were thought to apply to galactic systems composed of stars in rotational motion. The formula (\ref{28c}) involves $M(r)$, the mass within a sphere of radius $r$ and gives the velocity at the surface of that sphere but it says nothing about the way the mass is arranged in that sphere except that the centre of mass has to be at the centre. Thus $M(r) = M(r_i)=$ a constant for all $r> r_i$. Or alternatively expressed from (\ref{28c}), $v_2(r)$ is inversely proportional to $r^{1/2}$ for $r> r_i$. We can use this to write down a result to be used later. 
In the case when the radial variable $r$ is outside the region where the density fuction is non zero the mass function can be written as
\begin{eqnarray} 
M(r,r_i)=  M(r_i,r_i)=\  a\ constant.\label{28c1}
\end{eqnarray}
Let us now consider the case for $r\le r_i^\prime $. That is when the variable $r$ is within the region in which the density function $\rho (r_i^\prime)$ is constant and {\it not zero\/}. In that situation we can write  
\begin{eqnarray}
v^2_2 (r) =  \frac{G M(r,r_i^\prime)}{r}= \frac{G V(r) \rho(r,r_i^\prime)}{r}= \frac{4 \pi G r^2 \rho(r_i^\prime,r_i^\prime)}{3} \label{29c}
\end{eqnarray}
and because $\rho(r,r_i^\prime)$ is constant and equal to $\rho(r_i^\prime,r_i^\prime)$  for $r\le r_i^\prime$, we can now claim that $v_2(r)$ is directly proportional to $r$.
We can use this to write down a second result to be used later. 
In the case when the radial variable $r$ is inside the region where the density fuction is non zero the mass function can be written as
\begin{eqnarray} 
M(r,r_i^\prime)&=&  V(r)\rho(r,r_i^\prime).\label{29c1}\\
v^2_2 (r) &=&  \frac{G M(r,r_i^\prime)}{r}= \frac{G V(r) \rho(r,r_i^\prime)}{r}= \frac{4 \pi G r^2 \rho_{r_i^\prime} }{3}\label{311c}
\end{eqnarray}
where $M(r,r_i^\prime)$ is the total amount of pressure related mass within the sphere of radius $r$ and because $\rho(r,r_i^\prime)= \rho_{r_i^\prime} $, a constant  by (\ref{13c12}).
The two results (\ref{28c}) and (\ref{311c}) are the well know classical results from Newtonian theory the first giving the case when gravitational effects outside the source distribution are considered and the second case when gravitational effects within the source gravitational distribution are considered. The first case is clearly a good account of the observed speeds of planets in the solar system where the sun has the dominant effect on the planets.  The second case is the usually assumed form for how gravity would affect the motion of a free particle within a cavity inside the earth's interior or any other planet for that matter. The dark matter problem has arisen from astronomical observational measurements that indicate that the transverse velocities for stars at the edges of galaxies obey neither of the two cases above but rather obey an approximately constant relation between velocity $v_2(r)$ and the radial distance from the centre $r$. That is to say the measured galactic velocities squared  are greater than the theoretical value suggested by the tailing off first case above and less than the quadratic increasing second case above might imply. Clearly the formula (\ref{28c}) implies that if rotation velocities at some distance are to be higher then more mass within the region is required. Thus if theory is to adapt to observation the extra mass within the region has to be arranged in some special way. This is the {\it dark matter problem\/} of how much extra mass there has to be and where this missing mass needs to be located within the galactic volume so that it leads to constant velocity against distance rotation curves for cycling stars. There is nowadays some consensus that the missing or dark matter part of galactic structure is about four or five times the normal mass part. There seems to be quit a lot of variability over this estimate, it partly depend on how much dark energy the universe is assumed to contain at any time.  In the dust universe model, there is $75\%$ dark energy mass and $25\%$ normally gravitating matter present now. This seems to me to favour four for the ratio of dark matter to ordinary matter in the positively gravitating sector, $M_P^+$. These figures come from reference \cite{01:kmo} The analysis above of the rotation curves structure from (\ref{26c}) onwards has used the classical Newtonian theory and as we have seen that does not seem to explain why the galactic rotation curves are flat.  However, we have above the more elaborate general relativistic generalisation of Newtonian inverse square law gravitation theory that will be applied to the study of galactic structure in the next section.   
\section{General Relativistic Rotation Curves}
\setcounter{equation}{0}
\label{sec-grrc}
Let us now consider what the general relativistic generalisation of Newton's inverse square law can contribute to the problem of the missing mass and the rotation curves. In the general case for any $r$ we have,
\begin{eqnarray}
v^2_2(r) &=&  \frac{G}{r} \left(M^+( r, r_i)  + M_P(r, r_i^\prime)\right) -\frac{G M_\Lambda(r)}{r}   \label{30c}\\
M_P(r, r_i^\prime)&=& V(r) \frac{3 P(r_i ^\prime)}{c^2} \label{31c}\\
M_\Lambda(r) &=& V(r) \rho_\Lambda^\dagger. \label{32c}
\end{eqnarray}
Equation (\ref{30c}) is the general relativity rotation curve for transverse velocity in terms of distance from the origin $r$. The two following masses are the pressure induced mass and the dark energy mass within the spheres of radius $r$. These are clearly additional masses within the region concerned that help determine the form of the function $v_2(r) $. Pressures in cosmology are used in conjunction with an {\it equation\/} of state that expresses the pressure in term of a related density and a dimensionless function denoted by $\omega $ that can depend on space and time parameters as below but with no time dependence and only radial spatial distance involved for this problem.
\begin{eqnarray}
P(r, r_i^\prime)&=& c^2\rho (r, r_i^\prime)\omega (r, r_i^\prime) \label{33c}\\
M_{GR}^+(r)&=&M^+( r, r_i)+ M_P(r, r_i^\prime) = V(r)\rho(r,r_i) + V(r) \frac{3 P(r,r_i ^\prime)}{c^2} \label{34c}\\
&=& V(r)  \rho(r,r_i) + V(r) 3\rho (r, r_i^\prime) \omega (r, r_i^\prime). \label{35c}
\end{eqnarray}
It is not obvious what the function $\omega (r, r_i^\prime)$ should be except that we know that it represents positively gravitating material. In the case of negatively gravitating material associated with the $\Lambda$ term,  we do know it has the value $-1$. Thus a first reasonable shot at the value for the $\omega$ above is the value $+1$ which gives for equation (\ref{35c}) 
\begin{eqnarray}
M_{GR}^+= V(r)( \rho(r,r_i) +  3\rho (r, r_i^\prime)). \label{36c}
\end{eqnarray}
At equation (\ref{36c}) we see just how much, $3\rho (r, r_i^\prime))$, additional mass the pressure term from general relativity adds to the classical density term $\rho(r,r_i)$. Further, $ r_i^\prime$ is a model adjustable parameter so that it can be chosen to have any suitable value. The two densities are also model adjustable so that, if we decide that greater than four to one is the correct ratio of pressure induced mass $M^+_P$ to normal mass $M^+_N$, then  $ \rho (r, r_i^\prime)$ can be chosen so that
\begin{eqnarray}
3\rho (r, r_i^\prime)=4\rho(r,r_i),\ r<r_i\label{37c}
\end{eqnarray}
with $r_i^\prime > r_i $. That is to say, the mass distribution and the pressure distribution differ in value by a factor $3/4$.
The general relativity rotation curve for this case, without taking into account the $\Lambda$ term would be
\begin{eqnarray}
v^2_2(r) &=&  \frac{4\pi r^2G}{3}  \left( \rho(r,r_i)   + 4\rho(r,r_i) \right) \label{38c}\\
\frac{ M_P^+}{ M^+_N }&=& \frac{3\rho(r,r_i^\prime) {r_i^\prime}^3 }{\rho( r,r_i) {r_i}^3}= 4\frac{{r_i^\prime}^3 }{{r_i}^3}\ >4,\label{39c}
\end{eqnarray}
where $ r$ is a radial position inside both mass distributions and at (\ref{39c}), we have obtained the ratio of halo mass to visible mass to be greater than $4$.
It is clearly easy to find the values for the quantities concerned, the $\rho(r,r_i)$s and the $r_i$s to find any value that might be determined from experiment to be the correct value Thus I have constructed a simple computer program using mathematica to display graphically a whole range of rotation curves in the general relativity case that correspond to the astronomically observed rotation curves. Diagrams $1,2$ and $3$ from this program can be found on page $12$ of \href{http://www.maths.qmul.ac.uk/~jgg/gil127.pdf}{gil127}. 
This program in Mathematica note book language can be downloaded in the file $grcs.nb$ from my website at \href{http://www.maths.qmul.ac.uk/~jgg/grcs.nb}{QMUL Maths}.

I have carried through this case for identifying dark matter as originating from the pressure term that appears in Einstein's field equations and consequently also in the Friedman equations with an extremely simple model for a galaxy. The model is unrealistic in a number of ways, two constant spherically constant mass distributions have been used both of which terminate sharply at their radial limits $r_i$ and $r_i^\prime$. Galaxies are certainly not like that, particularly in the respect that the distribution of stars in the visible part of a galaxy tail off at a very indefinite outer boundary. Also the actual visible distributions observed are not uniformly constant but often have very complicated spiral shapes for example. One consequence of this sharp boundary aspect is that the curves I have calculated have sharp cusps so that the orbits calculated have to refer to stars at the lower boundary and indeed I have taken the constant curves to start at that boundary. I have used this simple model to avoid mathematical complications and to get quickly to a clear conclusion. Certainly the model could be made very much more realistic by giving the whole structure more sections with more character. However, in spite of the simplifications, I think it is established that Einstein's pressure term does account for the missing dark matter and the extra gravitational power that exists within a galaxy and holds it together.
\section{Conclusions Appendix 2}
\setcounter{equation}{0}
\label{sec-con2}
The dark matter problem has been around for about a century. Fritz Zwicky\cite{69:zwi} was an early astronomer to remark on this problem following his studies of the masses of galaxy clusters. There has been vast numbers of papers written on this subject and many varied attempts to find the solution of this problem which is essentially the problem of explaining and locating what appears to be vast quantities of mass in the universe that appears to exert gravitational attraction but cannot be seen with most of the observation astronomical equipment that exists at present. One dominate attempt at solving this problem has been the suggestion that Newton's theory of gravity needs to be modified particularly in the way it describes gravity at large distances from the gravitational source and this has seemed to imply that Einstein's general theory of relativity would also need to be modified. This connection arises because Newton's theory of gravity is a limiting case of Einstein's Theory. However, there seems to have been little clear recognition of the actual {\it form and structure\/} taken by the Einstein general relativity replacement for the Newton inverse square law. This failure I have rectified in this and earlier papers by deriving and displaying the full content of the general relativity inverse square law of gravity (\ref{11c}). This formula contains the essential addition of Einstein's {\it dark energy mass\/} and its pressure contribution both contained in the mass term, $M^-_\Lambda$ and also the full positively gravitating mass density contribution which includes the pressure induced part both contained in the mass term, $M_P^+$.  The addition of Einstein's mass {\it pressure term\/} is here used to described the so called mysterious {\it dark matter\/}, apparently invisible  contribution, here identified as the halo, that explains the missing gravitating mass of galaxies or their clusters. I feel that this claim is reinforced by an observation I made in an earlier paper, 
\href{http://www.maths.qmul.ac.uk/~jgg/gil108.pdf}{QMUL Maths},
with regard to {\it dark energy\/} is also applicable in part to {\it dark matter\/} repeated here as follows:- Thus the mystery of the origin of the dark energy density, $ \rho_\Lambda = \Lambda c^2/(8\pi G)$ in Einstein's form or in my revised form $\rho^\dagger_\Lambda=2\rho_\Lambda $, within the universe is completely resolved by this theory. Possibly this is the reason that dark energy is not visible. It could be because $pressures$ are not usually visible and the {\it pressure status\/} of the dark energy density is its dominant characteristic. However, it seems to me that dark energy with approximately an equivalent density of $5$ hydrogen atoms per cubic meter would not be visible anyway.
\vskip 0.5cm
\leftline{\bf Appendix 3: Mass of the Universe}
\vskip 0.5cm
\centerline{July 12, 2010}
\vskip 0.5cm  
\section{Abstract Appendix 3}
A theoretical value for the total positively gravitating mass of the universe is implied by the mathematical structure of the dust universe model. A simple formula is obtained that gives the value of this mass quantity in terms of Newton's gravitational constant, $G$, the Cosmological constant or Einstein's Lambda, $\Lambda$, and the velocity of light, $c$. This result depends on taking a fundamental view of an epoch time conditioned relation, obtained earlier, between the universe's content of positively gravitating mass density and the universe's content of negatively gravitating mass density, $\rho_\Lambda^\dagger=2\rho_\Lambda$, where the last quantity mentioned is Einstein's dark energy density. The value obtained is approximately $2.00789 \times 10^{53}\ kg$, The approximation aspect depends on the currently measured or assumed values for $G$ and $\Lambda$. 
\vskip 0.2cm 
\centerline{Keywords: Cosmology, Dust Universe, Dark Energy, Dark matter}
\centerline{Cosmological Constant, Friedman Equations}
\centerline{General Relativity, Newton's Gravitation constant}
\vskip 0.3cm
\centerline{PACS Nos.: 98.80.-k, 98.80.Es, 98.80.Jk, 98.80.Qc}
\section{Introduction Appendix 3: Mass of the Universe}
\setcounter{equation}{0}
\label{sec-intro3}
In the following pages, I shall demonstrate that the dust universe model can be used to arrive at a formula for the total positively gravitating mass of the universe. This is achieved with the help of a time conditioned relation between positively gravitating mass and the negatively gravitating mass now thought to pervade the whole universe and described by Einstein with his cosmological constant $\Lambda$. The following theory structure from the dust universe model is required for this project. The density functions for positively gravitating mass, dark energy and the ratio, $r_{\Lambda,DM} (t)$, of dark energy to positively gravitating mass $\rho(t)$, as functions of time are respectively represented by
\begin{eqnarray}
\rho (t) &=& (3/(8\pi G))(c/R_\Lambda)^2\sinh^{-2}(3 c t/(2R_\Lambda))\label{e1}\\
\rho^\dagger_\Lambda &=& (3/(4\pi G))(c/R_\Lambda)^2\label{e2}\\
r_{\Lambda,DM} (t) &=&\rho^\dagger_\Lambda /\rho (t) = 2 \sinh^{2}(3 c t/(2R_\Lambda)) \label{e3}\\ 
r_{\Lambda,DM} (t_c) &=& 2 \sinh^{2}(3 c t_c/(2R_\Lambda))= 1\label{e4}\\
t_c&=& (2 R_\Lambda/(3 c)) \coth ^{-1}(3^{1/2})\approx 2.1723367 \times 10^{17}\ s.\label{e4.1}
\end{eqnarray}
The time $t_c$ at which the acceleration changes from negative to positive is given by equation (\ref{e4.1}) above. This time is a fundamental constant which only depends on $\Lambda$ and $c$ and notably does not depend on mass.
The two equations (\ref{e1}) and (\ref{e2}) which together imply (\ref{e3}), I now see as spatially {\it local\/} fundamental properties of space time in the dust universe model. All the structure of this model can be obtained form these two equations by at least two different interpretations of the positively gravitating mass density function $\rho (t)$. The original interpretation of this density in the full universe context is to associate it with a volume, $V_U (t) $, and an {\it arbitrary input constant mass\/} quantity, $M_U$, for the universe of the form
\begin{eqnarray}
\rho (t)&=&M_U/V_U(r(t)) \label{e5}\\
V_U(r(t))&=& \frac{4\pi}{3} r^3(t) \label{e6}
\end{eqnarray}
so making it possible to locate the form and dependence on time of the radius, $r(t)$, of the expanding or contracting universe. This is the sense that I now see the first two equations as fundamental. Given next is the structure for the radius of the universe $r(t)$ that can be deduced from above for the positive time branch of the theory.
\begin{eqnarray}
r(t) & = & (R_\Lambda /c)^{2/3} C^{1/3} \sinh^{2/3} (3ct/(2R_\Lambda)) \label{e6.1}\\
b & = &  (R_\Lambda /c)^{2/3} C^{1/3}\label{e6.2}\\
C &=& 8\pi G \rho (t) r^3/3=2M_UG\label{e6.3}\\
R_{\Lambda} &=& |3/\Lambda|^{1/2}\label{e6.4}\\
r(t)&=& b\sinh^{2/3} (3ct/(2R_\Lambda))\label{e6.5}\\
r^3(t_c)&=& (R_\Lambda/c)^2M_UG.\label{e6.6}
 \end{eqnarray}
Firstly, let us see how far the two equations (\ref{e1}) and (\ref{e2}) can be interpreted regarding them as fundamental. The expression for my version of dark energy density, (\ref{e2}), which is in fact twice Einstein's version, presents directly no physical interpretational clue. In fact both versions are equally puzzling. However, if we invoke the dimensional structure of the gravitation constant, $G$,
\begin{eqnarray}
G=M_G^{-1} L_G^3 T_G^{-2}\label{e7}
\end{eqnarray}
expressing $G$ in terms of suitably powered mass, length, and time constant parameters giving the value of $G$, it is clear that two of the parameters can be chosen at random leaving the third to be determined by the expression (\ref{e7}). Thus let us make the choices $L_G=R_\Lambda$ and $T_G =R_\Lambda/c$ leaving $M_G$ to be determined by the original expression for $G$,
\begin{eqnarray}
G&=&M_G^{-1} R_\Lambda^3 (R_\Lambda/c )^{-2})\label{e8}\\
&=& M_G^{-1} R_\Lambda c^{2}\label{e9}\\
M_G &=& \frac{R_\Lambda c^{2}}{G} \label{e10}\\
r^{3}(t_c)&=&(R_\Lambda)^3\frac{M_U}{M_G}.\label{e10.1}
\end{eqnarray}
The last but one entry above gives the consequent value for $M_G$ necessary to ensure that $G$ remains invariant in value under these substitutions. The last entry gives the value of $r^3(t_c)$, (\ref{e6.6}), in terms of the new parameters. In the next display, I give the form that Einstein's version of the dark energy density takes followed by the form that my version of the dark energy density takes in terms of the substituted parameters
\begin{eqnarray}
\rho_\Lambda &=&  M_G /(8\pi R_\Lambda^3/3)\label{e11}\\
\rho^\dagger_\Lambda &=&  M_G /(4\pi  R_\Lambda^3/3).\label{e12}
\end{eqnarray}
There is no issue here as to which of these two versions is correct. They are both correct but represent different aspects of the dark energy spatial material.
Einstein's version is a raw energy density describing physically his mathematical $\Lambda$ term from the stress energy momentum tensor in his field equations. My version is the pressure enhanced version required by the stress energy momentum tensor to fully describe the gravitational effect of dark energy and gives the value of positive energy density (\ref{e12})  for effective negatively gravitational mass everywhere and for all time. Einstein's version (\ref{e11})  is correctly half this value. Thus the physical meaning for both of the versions is clearly represented in terms of the new parameters. The expression (\ref{e12})
states that the effective physical dark energy density is equal to having a mass quantity $M_G$ enclosed in a desitter volume of size $V_\Lambda=4\pi R_\Lambda^3/3$. The formula (\ref{e12}) for $\rho^\dagger_\Lambda $ is a very definite and simple result giving information about the universe's astro-space character and our knowledge of its value is only restricted by the accuracy with which we know by measurement or otherwise the constants $G$ and $\Lambda$.
From the interpretation of the dust universe model using (\ref{e5}) and (\ref{e6}) together with an arbitrary value of the positively gravitating mass within the universe boundary, $M_U$, say, let us the consider the formula (\ref{e3}) taken at epoch time, $t=t_c$, the time when the universe has zero radial acceleration. At that time, the formula reduces to
\begin{eqnarray}
M_U/V_U(r(t_c)) &=&\rho (t_c)= \rho^\dagger_\Lambda= M_G / (4\pi  R_\Lambda^3/3) \label{e13}\\
&=& M_U/(4\pi R_\Lambda^3M_U/(3M_G))=M_G/(4\pi  R_\Lambda^3/3).\label{e14}
\end{eqnarray}
The last equality in (\ref{e13}) giving the alternatively expressed density obtained earlier and because $ r^3(t_c) =R_\Lambda^3M_U/M_G$ by (\ref{e10.1}). Taking into account both equations we see that a change in the value of the mass of the universe in the formulae can be made consistently provided the change in value of $r(t_c)$ that is involved in the change of parameters is taken into account. However, it is also clear that taking the definite and fixed value $M_G$ for the mass of the universe in place of the original $M_U$, some what arbitrary though informed choice, does make some simplification. In particular $r(t_c) = R_\Lambda$ is the result.
Using the same sequence of steps in the case of a primitive galaxy, which is just an un clumped region of definite amount of mass $M_g$ in a volume $V_g$ at time $t_c$, say, we get 
 \begin{eqnarray}
M_g/V_g(r(t_c)) &=&\rho (t_c)= \rho^\dagger_\Lambda= M_G / (4\pi  R_\Lambda^3/3) \label{e15}\\
&=& M_g/V_g(R_{\Lambda,g})=M_g/(4\pi  R_{\Lambda,g}^3/3).\label{e16}
\end{eqnarray}
Taking into account both equations above, in this case, because $r(t_c)$ equals a smaller radius, $ R_{\Lambda,g}$, it follows that
\begin{eqnarray}
M_G / (4\pi  R_\Lambda^3/3) &=& M_g/(4\pi  R_{\Lambda,g}^3/3)\label{e17}\\
&\implies&  \frac{M_g}{ M_G }= \left(\frac{ R_{\Lambda,g}}{ R_\Lambda}\right)^{3}, \label{e18} 
\end{eqnarray}
which is the sensible statement that the ratio of the mass of a galaxy  to the mass of the universe is equal to the ratio of the volume of a  galaxy to the volume of the universe at time $t_c$. Thus the amount of mass used in the formalism is arbitrary and the formalism gives consistent results provided account is taken of all relevant facets.
\section{Conclusions Appendix 3}
\setcounter{equation}{0}
\label{sec-con4}
The dust universe model can be used to calculate a theoretical value for the mass of the universe. This quantity of mass is represented above as
\begin{eqnarray}
M_G &=&\frac{ R_\Lambda c^{2}}{G}= \left(\frac{3c^4}{\Lambda G^2}\right)^{1/2} \approx 2.00789 \times 10^{53}\ kg.\label{e19}
\end{eqnarray}
I cannot claim that the value of $M_G$ {\it is\/} the mass of the universe. Any very large quantity of mass will work in the formalism because the formalism can describes any packet of cosmologically conserved region of {\it substratum\/} mass over epoch time. However, the quantity of mass $M_G$ is quite definite in value and it does give a clear physical explanation of the dark energy density $\rho_\Lambda^\dagger$, picking up the physically significant version of this quantity, rather than the less physically significant Einstein version. Another thing in its favour is that it confers on the radius of the universe at time, $t_c$, a definite and special status, $r(t_c)=R_\Lambda$, which goes well with the special and invariant status of the time $t_c$ itself. $M_G$ is about a power of ten larger than some estimates I have seen. Of course, I would like it to turn out to be the actual mass of the universe. Only time will tell.
\vskip 0.5cm
\leftline{\bf Acknowledgements}
\vskip 0.5cm
\leftline{I am greatly indebted to Professors Clive Kilmister and} 
\leftline{Wolfgang Rindler for help, encouragement and inspiration}

\end{document}